\documentclass[letter]{jpsj2} 
\usepackage{color}
\usepackage{ulem}

\begin{document}
\title{Iridium Orbital Crossover at the Structural Phase Transition in Ca$_{10}$(Ir$_4$As$_8$)(Fe$_{2-x}$Ir$_x$As$_2$)$_5$}

\author{
Kento \textsc{Sugawara}$^{1}$,
Naoyuki \textsc{Katayama}$^{1}$\thanks{E-mail address: katayama@mcr.nuap.nagoya-u.ac.jp},
Yuki \textsc{Sugiyama}$^{1}$,
Takafumi \textsc{Higuchi}$^{1}$,
Kazutaka \textsc{Kudo}$^{2}$,
Daisuke \textsc{Mitsuoka}$^{2}$,
Minoru \textsc{Nohara}$^{2}$, and Hiroshi \textsc{Sawa}$^{1}$
}

\inst{
$^{1}$Department of Applied Physics, Nagoya University, Nagoya 464-8603, Japan\\
$^{2}$Department of Physics, Okayama University, Okayama 700-8530, Japan
}

\abst{
We report a structural transition found in Ca$_{10}$(Ir$_4$As$_8$)(Fe$_{2-x}$Ir$_x$As$_2$)$_5$, which exhibits superconductivity at 16 K, with a layer of divalent iridium coordinated by arsenic in between Fe$_2$As$_2$ layers. The $c$-axis parameter is doubled below a structural transition temperature of approximately 100 K, while the tetragonal symmetry with space group $P4/n$ (No. 85) is unchanged at all temperatures measured. Our synchrotron x-ray diffraction study clearly shows displacements along the $z$-direction occur in half of the iridium sites, resulting in a complex orbital ordering pattern. Combining our theoretical calculation of the 5$d$ orbital energies with structural data, we propose the iridium ``orbital crossover" transition between the $d_{xy}$ and $d_{z^2}$ orbitals.}

\kword{Iron-based superconductors, orbital crossover, Ca$_{10}$(Ir$_4$As$_8$)(Fe$_{2-x}$Ir$_x$As$_2$)$_5$}

\maketitle

Since the discovery of superconductivity in LaFeAsO$_{1-x}$F$_x$ \cite{rf:1}, the high $T_c$ superconducting mechanism has been attributed to both the magnetic and structural properties of the material. All iron based superconducting families identified so far consist of the same structural motif of Fe$_2$As$_2$ layers and spacer layers \cite{rf:2, rf:4, rf:3, rf:5, rf:6, rf:7, rf:8, rf:9, rf:10, rf:11,rf: 12, rf:13, rf:14, rf:15, rf:16, rf:17, rf:19, rf:20}. Therefore, the crystal structure of a material can be classified according to the spacer layer. Examples include (i) 1111-type structure with slabs of rare-earth oxides or alkali-earth fluorides with a fluorite-type structure \cite{rf:1, rf:7}, (ii) 111- and 122-type structure with alkali or alkali-earth ions \cite{rf:8, rf:6}, (iii) 32522-type structure and its derivative with complex metal oxides with perovskite-type structure or combinations of perovskite- and rocksalt-type structures \cite{rf:9, rf:10, rf:11,rf: 12, rf:13, rf:14, rf:15, rf:16}, (iv) 112-type structure with arsenic zigzag chains \cite{rf:2, rf:4}, and (v) 10-4-8 and 10-3-8 phases as in Ca$_{10}$(Pt$_n$As$_8$)(Fe$_{2-x}$Pt$_x$As$_2$)$_5$ with $n$ = 3 and 4 and Ca$_{10}$(Ir$_4$As$_8$)(Fe$_{2-x}$Ir$_x$As$_2$)$_5$ \cite{rf:3, rf:5, rf:19, rf:20}. Usually, spacer layers are insulating in nature without any degrees of freedom. If there exist degree of freedom of the spacer layers, such as charge, orbital and lattice, we expect that their orderings or fluctuations should affect superconductivity in the Fe$_2$As$_2$ layers through the intimate interlayer coupling, leading to a new controllable parameter.

\begin{figure}[t]
\begin{center}
\includegraphics[width=7.6cm]{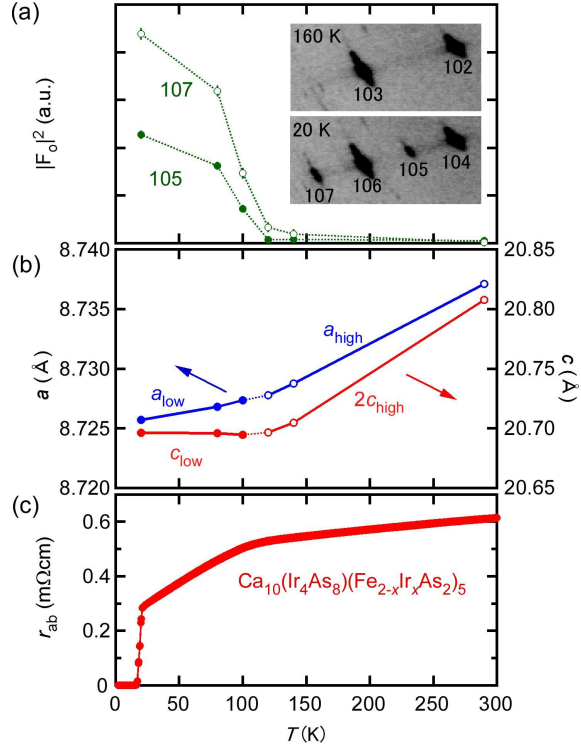}
\caption{\label{fig:Fig1}
(Color online)  (a) Temperature dependence of the square of the observed structure factor, $|F_o|^2$, which is proportional to the superstructure peak intensity. (b) Lattice parameters at each temperature, where $c_{low}$ and $c_{high}$ indicate the $c$-axis parameters in the low-temperature and high-temperature phases, respectively. (c) Temperature dependence of the electrical resistivity parallel to the $ab$-plane, $\rho_{\rm ab}$, for Ca$_{10}$(Ir$_4$As$_8$)(Fe$_{2-x}$Ir$_x$As$_2$)$_5$. See ref.[15] for more details. Note that a small kink appears near 100 K. }
\end{center}
\end{figure}

In order to address this issue, we present a structural study on Ca$_{10}$(Ir$_4$As$_8$)(Fe$_{2-x}$Ir$_x$As$_2$)$_5$ with $T_c$ of 16 K. This compound was first reported by Kudo $et~al.$ with the chemical formula of Ca$_{10}$(Ir$_4$As$_8$)(Fe$_2$As$_2$)$_5$ based on the synchrotron x-ray diffraction analysis \cite{rf:3}. In this study, we re-analyzed the chemical composition and found a small amount of Ir substitution for Fe, which will be discussed later. Both Ca$_{10}$(Ir$_4$As$_8$)(Fe$_{2-x}$Ir$_x$As$_2$)$_5$ and a platinum derivative, Ca$_{10}$(Pt$_4$As$_8$)(Fe$_{2-x}$Pt$_x$As$_2$)$_5$, reported by Ni $et~al.$, crystallize in tetragonal structures with the space group $P4/n$ \cite{rf:3, rf:5}. The most significant difference between them is related to the electron configurations; Pt$^{2+}$ (5$d^8$) forms a closed-shell configuration with a completely filled $d_{xy}$ orbital in the square-planar coordination, whereas $d_{xy}$ of Ir$^{2+}$ (5$d^7$) is formally half-filled, resulting in a metallic nature. The two materials display different temperature dependence in electrical conductivity, which is likely due to the different electron configurations of the spacer layers. The resistivity of Ca$_{10}$(Pt$_4$As$_8$)(Fe$_{2-x}$Pt$_x$As$_2$)$_5$ decreases linearly with temperature \cite{rf:5}, whereas the resistivity of Ca$_{10}$(Ir$_4$As$_8$)(Fe$_{2-x}$Ir$_x$As$_2$)$_5$ exhibits an unusual kink near 100 K, as shown in Fig.\ref{fig:Fig1}(c), suggesting an unusual transition occurs that is related to the electron configuration of the divalent iridium.

In this letter, we present a synchrotron x-ray structural study on Ca$_{10}$(Ir$_4$As$_8$)(Fe$_{2-x}$Ir$_x$As$_2$)$_5$. Our analysis clearly exhibits the appearance of superstructure peaks below 100 K, indicating a doubled period along the $c$-axis. In the low-temperature phase, half of the iridium ions shift along the $c$-axis, displacing the surrounding arsenic ions. Together with theoretical calculations based on the point-charge model, we conclude that the transition is an orbital crossover transition between 5$d_{xy}$ and 5$d_{z^2}$ of the divalent iridium.

\begin{table}[t]
\begin{center}
 \caption{Data collection and refinement statistics for the synchrotron X-ray structure determination of Ca$_{10}$(Ir$_4$As$_8$)(Fe$_{2-x}$Ir$_x$As$_2$)$_5$.}
 \label{tab:table1}
 \vspace{3mm}
{\tabcolsep=3mm
 \begin{tabular}{ll}
  \hline
  \hline
Ca$_{10}$(Ir$_4$As$_8$)(Fe$_{2-x}$Ir$_x$As$_2$)$_5$ &  20 K\\
  \hline
  \hline
 \multicolumn{2}{c}{Data Collection }\\
 Crystal System & tetragonal \\
 Space Group & $P4/n$\\
 $a$ (\AA) & 8.7236(17) \\
 $c$ (\AA) & 20.6799(18)  \\
 $R_{\rm merge}$ & 0.0614\\
 $I$ / $\sigma$ $I$ & $>$2\\
 &\\
 \multicolumn{2}{c}{Refinement }\\
 Resolution (\AA) &  $>$0.50\\
 No. of Unique Reflections & 3457\\
 $R$1  & 0.0409\\
 \hline
  \hline
 \end{tabular}}
\end{center}
\end{table}

\begin{table}[hbt]
\renewcommand{\arraystretch}{1}
\caption{
Structural parameters of Ca$_{10}$(Ir$_4$As$_8$)(Fe$_{2-x}$Ir$_x$As$_2$)$_5$ with the space group $P4/n$ at 20 K\cite{rf:21}. 
}
\label{tab:table2}
\begin{tabular}{lccccc}
\hline
\hline
\multicolumn{5}{c}{Ca$_{10}$(Ir$_4$As$_8$)(Fe$_{2-x}$Ir$_x$As$_2$)$_5$ at 20 K.}\\
\hline
\multicolumn{5}{c}{Atomic Position}\\
Site~~  & Occ.~~ & $x/a$ & $y/b$ & $z/c$ \\
\hline
Ir(1)~~ & 1 & 1/4 & -1/4 & 0.247974(14) \\
Ir(2a)~~ & 1 & -1/4 & -1/4 & 0.20958(2) \\ 
Ir(2b)~~ & 1 & 1/4 & 1/4 & 0.27449(2) \\ 
Fe(1c)~~ & 0.911(3) & -3/4 & -1/4 & 0 \\
Ir(1c)~~ & 0.089 & -3/4 & -1/4 & 0 \\
Fe(2c)~~ & 0.938(2) & -0.34874(9) & -0.44598(9) & 0.00378(4) \\ 
Ir(2c)~~ & 0.062 & -0.34874 & -0.44598 & 0.00378 \\
Fe(1d)~~ & 0.911(3) & 1/4 & -1/4 & 1/2 \\
Ir(1d)~~ & 0.089 & 1/4 & -1/4 & 1/2 \\
Fe(2d)~~ & 0.938(2) & -0.05035(9) & -0.35010(9) & 0.50078(4) \\
Ir(2d)~~ & 0.062 &  -0.05035 & -0.35010 & 0.50078 \\
As(1a)~~ & 1 & -0.14200(8) & -0.00692(8) & 0.25072(3) \\ 
As(2a)~~ & 1 & -1/4 & -1/4 & 0.07647(6) \\
As(3a)~~ & 1 & -0.54546(8) & -0.34881(8) & -0.06679(3) \\ 
As(1b)~~ & 1 & 0.13935(8) & 0.00283(8) & 0.24954(3) \\
As(2b)~~ & 1 &  -1/4 & -1/4 & 0.56830(6) \\
As(3b)~~ & 1 & 0.04737(7) & -0.15073(7) & 0.43176(3) \\ 
Ca(1a)~~ & 1 & 0.05329(14) & -0.15489(14) & 0.12963(6) \\ 
Ca(2a)~~ & 1 & 1/4 & 1/4 & 0.13355(11) \\
Ca(1b)~~ & 1 & -0.04915(14) & 0.15412(15) & 0.36948(6) \\
Ca(2b)~~ & 1 & -1/4 & -1/4 & 0.36103(12) \\
\hline
\hline
\end{tabular}
\end{table}
%


Single crystals of Ca$_{10}$(Ir$_4$As$_8$)(Fe$_{2-x}$Ir$_x$As$_2$)$_5$ were grown at Okayama University. The synthesis details have already been reported \cite{rf:3}. The single-crystal x-ray diffraction experiments were performed at SPring-8 BL02B1 (Hyogo, Japan) using single crystals with typical dimensions of 40 $\times$ 40 $\times$ 70 $\mu$m$^3$. The x-ray wavelength was 0.52 \AA. The obtained lattice parameters and refined conditions are summarized in Table \ref{tab:table1}. The electrical resistivity $\rho_{\rm ab}$ (parallel to the $ab$-plane) measurements were performed using the standard DC four-terminal method in a physical property measurement system (Quantum Design PPMS).

In our single crystal x-ray diffraction experiments, superstructure peaks emerged near 100 K, which nearly correspond to the kink in the electric resistivity measurement, and the peaks increased as the sample was cooled, as shown in Fig.\ref{fig:Fig1}(a). The superstructure peaks indicate a doubled period along the $c$-axis direction below the transition. As shown in Fig.\ref{fig:Fig1}(b), lattice parameters show no discontinuous jump at 100 K, suggesting that the transition is second order.

Through careful investigation of the Laue symmetry and the extinction rule of $h$+$k$$\neq$2$n$ for $hk$0, we found that the structural symmetry at 20 K was tetragonal with the space group $P4/n$, indicating that the space group was unchanged across the transition. Our precise definition of the composition showed a small amount of Ir substituted for Fe in the present samples, which was not reported in the previous study \cite{rf:3}. The low-temperature structural data is summarized in Table \ref{tab:table2}. 

\begin{figure*}[t]
\begin{center}
\includegraphics[width=14.2cm]{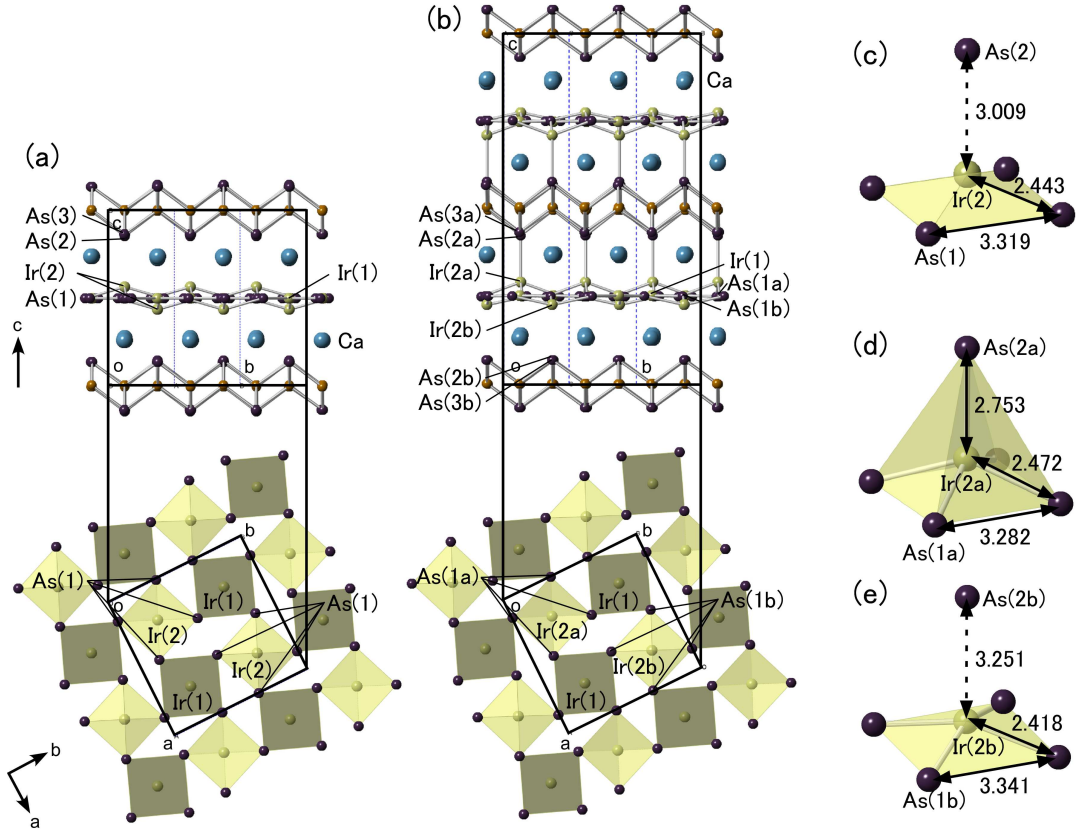}
\caption{\label{fig:Fig2}
(Color online) Crystal structures of Ca$_{10}$(Ir$_4$As$_8$)(Fe$_{2-x}$Ir$_x$As$_2$)$_5$ at (a) 290 K and (b) 20 K. The atomic distances between the iridium and the arsenic ligands, where (c) represents Ir(2) and the ligands at 290 K, while (d) and (e) show Ir(2a) and Ir(2b) with the arsenic ligands at 20 K, respectively. }    
\end{center}
\end{figure*}

The space group $P4/n$ is preserved at the lowest measured temperatures, while some atomic sites split into two crystallographically inequivalent sites, as shown in Fig.\ref{fig:Fig2}(a) and (b). The most notable structural features appear near the iridium ions. There are two iridium sites at high temperatures: Ir(1) is at a coplanar site and Ir(2), which splits into Ir(2a) and Ir(2b) in low temperatures, is at a non-coplanar site with respect to the As square. At the transition, Ir(2b) shifts toward the center of the As square, as shown in Fig.\ref{fig:Fig2}(c) and (e). In contrast, Ir(2a) is displaced toward the As(2a) ion in the adjacent Fe$_2$As$_2$ layer, resulting in an almost 10\% reduction in the Ir(2a)-As(2a) distance than that at 290 K, as shown in Fig.\ref{fig:Fig2}(c) and (d). Note that As(2a) also shifts toward Ir(2a) in low temperature, suggesting a strong tendency toward the bonding formation between Ir(2a) and As(2a). As a result of the displacement of iridium, the As(1b) square expands, whereas the As(1a) square contracts, as shown in Fig.\ref{fig:Fig2}(d) and (e).

\begin{figure}[t]
\begin{center}
\includegraphics[width=8.2cm]{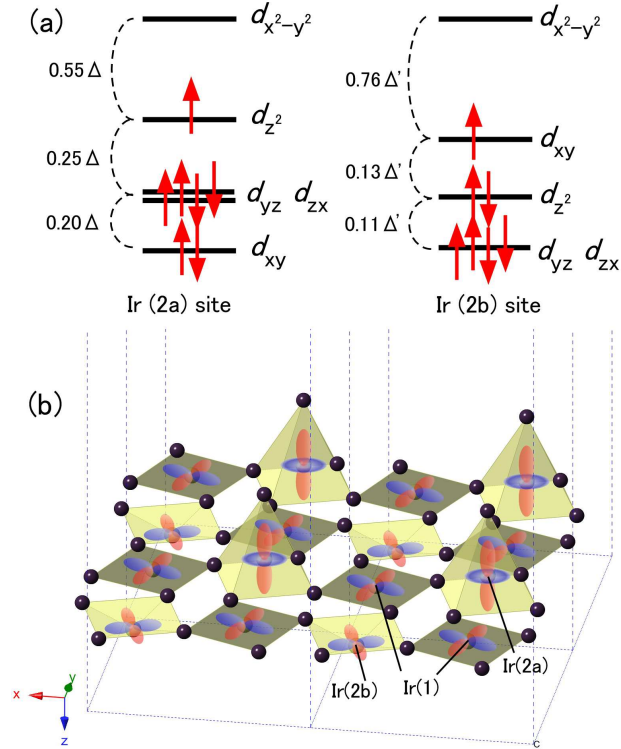}
\caption{\label{fig:Fig3}
(Color online) (a) Energy splitting of 5$d$ orbitals for Ir(2a) and Ir(2b) estimated from our calculation. $\Delta$ and $\Delta$' are proportional to crystal field splitting energies. (b) Schematic of the expected orbital ordering patterns of iridium at low temperatures.}
\end{center}
\end{figure}

The experimentally observed strong bonding tendency between Ir(2a) and As(2a) leads us to expect an unpaired electron in the Ir(2a) $d_{z^2}$ orbital, which spreads toward As(2a) directions. Based on the crystal field theory, this orbital exists in the square-pyramidal crystal field around the divalent Ir(2a) ion with a 5$d^7$ electronic state. Because the $d_{z^2}$ orbital has higher energy than the $d_{xy}$ orbital in the square-pyramidal crystal field, an unpaired electron may exist in the nondegenerate $d_{z^2}$ orbital. Here, the divalent electric state of iridium has been suggested experimentally using x-ray photoelectron spectroscopy (XPS) \cite{rf:3}. To confirm that the $d_{z^2}$ orbital energy is higher than the $d_{xy}$ orbital energy, we used a theoretical calculation based on the point-charge model by assuming the formal electron valences from the chemical formula: Ca$^{2+}_{10}$(Ir$^{2+}_4$(As$_2$)$^{4-}_4$)(Fe$^{2+}_{2-x}$Ir$^{2+}_x$As$^{3-}_2$)$_5$$\cdot$2e$^{2-}$. In this calculation, the Coulomb interaction between an unpaired electron in the Ir 5$d$ orbital and the surrounding As ions is given by

\begin{equation}
  v(\mbox{\boldmath $r$}) = - \sum \frac{Z_R e^2}{\epsilon_{o}|\mbox{\boldmath $R$}_i-\mbox{\boldmath $r$}|},
\end{equation}
where $\epsilon_{o}$ indicates the dielectric constant of vacuum, and $\mbox{\boldmath $R$}_i$ expresses the coordinate of the $ith$ As ion, which is deduced from our x-ray diffraction data (Table \ref{tab:table1}). Here, $Z_R$ represents the formal electron valence for each ion presented above: 2- for As(1a) and As(1b), and 3- for As(2a) and As(2b), respectively. The calculation demonstrated that the $d_{z^2}$ orbital energy was higher than the $d_{xy}$ orbital energy due to the Coulomb potential from the surrounding ions, as shown in Fig.\ref{fig:Fig3}(a), indicating that the coordination structure around the Ir(2a) ion can be regarded as  square-pyramidal, as shown in Fig.\ref{fig:Fig2}(d). Here, $d_{yz}$ and $d_{zx}$ orbital energies are higher than $d_{xy}$ orbital energy due to the strong Coulomb potential from the trivalent As(2a).

On the other hand, when we applied similar calculations to Ir(2b), which shifts toward the center of the As square in low temperatures, the results clearly show that the $d_{xy}$ orbital energy is higher than the $d_{z^2}$ orbital energy, as shown in Fig.\ref{fig:Fig3}(a), suggesting a square-planar-type crystal field exists around the Ir(2b) ion. A similar result can be obtained for the Ir(1) site as well, which is located at the coplanar atomic site with respect to the As square. As a result, the present system displays complex orbital ordering patterns for the divalent iridium ions, as shown in Fig.\ref{fig:Fig3}(b). Considering that an unpaired electron occupies the 5$d_{xy}$ orbital at high temperatures, as shown in the previous study \cite{rf:3}, our experimental results and theoretical calculations clearly demonstrate the transition between the $d_{xy}$ and $d_{z^2}$ orbitals of divalent iridium, which we can refer to it as an ``orbital crossover" transition of the divalent iridium. 

Because the degeneracy of the 5$d$ orbitals of Ir(2) is inherently lifted at high temperatures, the origin of the orbital crossover transition is unclear. The displacement of iridium may enhance the hybridization between the Ir 5$d$ and As 4$p$ orbitals, leading to a lowering of the total energy of the present system. As we mentioned above, the displacement of Ir(2a) causes a decrease in the Ir(2a)-As(2a) distances, suggesting a tendency toward hybridization between the Ir 5$d_{z^2}$ and As 4$p$ orbitals. Furthermore, the Ir(2b) ions with the unpaired electron in 5$d_{xy}$ orbital shifts toward the As square, enhancing the hybridization between the Ir 5$d_{xy}$ orbital and As 4$p$ orbitals. 

The commensurate periodicity between Fe$_2$As$_2$ and Ir$_4$As$_8$ layers may be responsible for the different 5$d$ orbital characters between Ir(2a) and Ir(2b) at low temperatures. In the present system, the unit cell contains four IrAs$_4$ squares per Ir$_4$As$_8$ layer, which is intimately coupled with the adjacent Fe$_2$As$_2$ layers. If both Ir(2a) and Ir(2b) display the same 5$d$ orbital characters, namely, displace toward the same directions with regard to the As squares, both As(1a) and As(1b) squares should expand or contract together associated with the displacement of iridium. This should lead to the destruction of the commensurate periodicity between Fe$_2$As$_2$ and Ir$_4$As$_8$ layers by an abrupt change in the size of the Ir$_4$As$_8$ planes across the transition. In the present system, the different 5$d$ orbital characters between Ir(2a) and Ir(2b) appear along with the expansion of the As(1b) squares and contraction of the As(1a) squares, which soften the change of the size of the Ir$_4$As$_8$ planes across the transition.


Such a lattice deformation can be viewed as an appearance of the ``breathing" square lattice. The similar breathing lattices can be also found in other systems, such as pyrochlore \cite{rf:22} and perovskite \cite{rf:23} systems. Although the origins differ among them, the corner sharing structures are common feature. The lattice instability caused by the electronic state of the divalent iridium, associated with the lattice condition in the present system, displays the anomalous breathing deformation accompanied by the iridium orbital crossover.

Finally, we discuss the superconductivity at low temperatures. All Fe$_2$As$_2$ layers are crystallographically equivalent at high temperatures and split into two crystallographically inequivalent layers below the transition: one layer includes As(2b) and the other layer contains As(2a), which is slightly displaced toward the Ir$_4$As$_8$ layer, as shown in Fig.\ref{fig:Fig3}(b). The displacement of the arsenic ions associated with the iridium displacements introduce notable differences in the As-Fe-As bond angles between the two Fe$_2$As$_2$ layers, which should affect $T_c$ \cite{rf:18}; the angle ranges from 105.58(4)$^{\circ}$ to 112.29(4)$^{\circ}$ for the layer containing As(2a) and from 107.14(4)$^{\circ}$ to 110.69(4)$^{\circ}$ for the layer including As(2b), respectively, which can be compared with the angle (109.47$^{\circ}$) of the regular tetrahedron. The derivative of the As-Fe-As bond angles between layers indicates that the ordering of any degrees of freedom in the spacer layer can be available as a controllable parameter for $T_c$. 

Unlike Ca$_{10}$(Ir$_4$As$_8$)(Fe$_{2-x}$Ir$_x$As$_2$)$_5$, the platinum derivative Ca$_{10}$(Pt$_4$As$_8$)(Fe$_{2-x}$Pt$_x$As$_2$)$_5$ with space group $P4/n$ exhibits no structural phase transitions in the temperature dependence of the electric resistivity. Because the platinum derivative shows superconductivity with a high transition temperature of 25 K \cite{rf:3, rf:5}, we can naively speculate that the structural transition suppresses superconductivity in the present system. This, however, may allow us to enhance the superconducting transition temperature by suppressing the structural transition using other methods such as chemical doping or the application of pressure, leading to a further study of the correlation between superconductivity and the orbital crossover transition. 

In summary, we studied the nature of the structural transition found in Ca$_{10}$(Ir$_4$As$_8$)(Fe$_{2-x}$Ir$_x$As$_2$)$_5$, which undergoes a superconducting transition at 16 K, using synchrotron x-ray diffraction experiments. Our x-ray diffraction results reveal the displacement of iridium and the associated arsenic displacement below the transition temperature. Combined with the theoretical calculations of the crystal field splitting using the point-charge model, we conclude that the structural transition can be interpreted as an orbital crossover transition between the $d_{xy}$ and $d_{z^2}$ orbitals of divalent iridium. For further theoretical studies, the spin-orbit effects should be considered because the spin-orbit coupling is likely stronger in the 5$d$ orbitals than in 3$d$ and 4$d$ elements.

\begin{acknowledgements}
The work at Nagoya University was supported by a Grant-in-Aid for Scientific Research (No. 23244074). The work at Okayama University was partially supported by a Grant-in-Aid for Scientific Research (C) (No. 25400372) from the Japan Society for the Promotion of Science (JSPS) and the Funding Program for World-Leading Innovative R\&D on Science and Technology (FIRST Program) from JSPS. Some of the magnetization measurements were performed at the Advanced Science Research Center at Okayama University. The synchrotron radiation experiments performed at BL02B1 in SPring-8 were supported by the Japan Synchrotron Radiation Research Institute (JASRI) (Proposal Nos. 2012A0083, 2012B0083, 2013A0083, and 2013B0083).
\end{acknowledgements}

\end{document}